\documentclass[11pt,a4paper]{article}

\usepackage{jheppub}
\usepackage{amsmath}
\usepackage{tensor}
\usepackage{scalerel}
\usepackage{amssymb}
\usepackage{bbold}
\usepackage{perpage}
\usepackage[numbers,sort&compress]{natbib}
\usepackage{graphicx}
\usepackage{setspace}
\usepackage[english]{babel}
\usepackage[utf8x]{inputenc}
\usepackage[dvipsnames]{xcolor}
\definecolor{royalpurple}{rgb}{0.47, 0.32, 0.66}
\usepackage{hyperref}
\hypersetup{
    colorlinks=true, %set true if you want colored links
    linktoc=page,     %set to all if you want both sections and subsections linked
    urlcolor=MidnightBlue,
    anchorcolor=royalpurple,
    citecolor=royalpurple,
    filecolor=royalpurple,
    linkcolor=royalpurple,
    menucolor=royalpurple,
}

\MakePerPage{footnote}
\newcommand{\imunu}[0]{\indices{_\mu_\nu}}
\newcommand{\iij}[0]{\indices{_i_j}}
\newcommand{\bkg}[1]{\tensor[^{\scaleto{(0)}{6pt}}]{#1}{}}
\newcommand{\prt}[1]{\tensor[^{\scaleto{(1)}{6pt}}]{#1}{}}
\newcommand{\gif}[1]{\tensor[^{\scaleto{(gi)}{6pt}}]{#1}{}}
\newcommand{\bkgt}[2]{\tensor[^{\scaleto{(0)}{6pt}}]{#1}{#2}}
\newcommand{\prtt}[2]{\tensor[^{\scaleto{(1)}{6pt}}]{#1}{#2}}
\newcommand{\git}[2]{\tensor[^{\scaleto{(gi)}{6pt}}]{#1}{#2}}
\newcommand{\partiali}[0]{\partial\indices{_i}}
\newcommand{\partialj}[0]{\partial\indices{_j}}
\newcommand{\partialui}[0]{\partial\indices{^i}}

\title{On the holographic domain-wall/cosmology correspondence and scalar one-point functions}

\author[a]{Jean-Fran\c{c}ois Vaduret}

\affiliation[a]{Institutionen f\"{o}r fysik och astronomi, Uppsala Universitet,\\
Box 803, E-751 08 Uppsala, Sweden}

\emailAdd{jean-francois.vaduret.8245@student.uu.se}

\abstract{We review the construction of holography for cosmologies with an emphasis on asymptotically de Sitter cosmologies. We do this by composition of the \emph{AdS/CFT} correspondence and domain-wall/cosmology (\emph{DW/C}) correspondence with the novel added benefit of keeping track of the \emph{DW/C} marker $\eta$. This allows us to build explicitly the holographic dual of the \emph{DW/C} correspondence. We then test this framework on a toy model: the \emph{GPPZ} flow. We find the scalar one-point function for this flow to be invariant under the domain-wall/cosmology correspondence. Moreover, our construction seems to agree with previous literature on holography for cosmologies.}

\begin{document}
	\maketitle

    \section{Introduction}

        %Fundamentally speaking, physics is about the study of our vicinity. For cosmologists, however, this vicinity is the entire universe. So it's only natural we try to understand everything we can about our universe. And as things stand in research, we need all the tools we have at our disposal. One of the most relatively recent ones is holography \cite{tHooft:1993dmi}. It relates any theory of quantum gravity to a corresponding gravity-free quantum field theory (\emph{QFT}) in one dimension less. In other words, there exists a dictionary which translates any event in a quantum gravity theory, to one in a quantum field theory, at the price of dimensional reduction. Such a tool should offer new powerful computational frameworks to compute cosmological observables.

Research in an attempt to ally cosmology and holography has already started a little over twenty years ago, with works like \cite{Witten:2001kn,Strominger:2001pn,Strominger:2001gp,Maldacena:2002vr}. Then more recently, a new way to build this framework was proposed in \cite{McFadden:2009fg,McFadden:2010na}. The first step was to construct holography for cosmologies and give predictions for observables that could be compared with data from observations. The theoretical framework backing this up rests on the assumption that the gauge/gravity duality still holds all while using the Domain-Wall/Cosmology (\emph{DW/C}) correspondence \cite{hep-th/0602260,hep-th/0610253,hep-th/0609056} to holographically compute observables. In this setup, the advantage lies in the fact that holography offers a strong/weak coupling duality. This means that where gravitational dynamics are strongly coupled, the corresponding quantum field theory is weakly coupled. This may alleviate difficulties encountered in the study of early-time cosmologies, the inflationary period being an obvious candidate. In fact, massive efforts were made in the recent literature to holographically describe inflationary cosmologies (see for example \cite{Maldacena:2011nz,Hartle:2012qb,Hartle:2012tv,Schalm:2012pi,Bzowski:2012ih,Mata:2012bx,Garriga:2013rpa,Ghosh:2014kba,Garriga:2014ema,Kundu:2014gxa,McFadden:2014nta,Arkani-Hamed:2015bza,Kundu:2015xta,Hertog:2015nia,Garriga:2015tea,Garriga:2016poh,Hawking:2017wrd,Arkani-Hamed:2018kmz,Adam:2019ayj,Achucarro:2018ngj,Garriga:2014fda}). While on the other hand, it is possible to constrain and test holographic cosmology by using data from observations as shown in \cite{Afshordi:2016dvb,Afshordi:2017ihr,Easther:2011wh,Dias:2011in}.

The holographic dictionary connecting domain-wall solutions and quantum field theories has been extensively studied and is believed to be well understood in some explicit realization of holography---\emph{e.g.} the \emph{AdS/CFT} correspondence \cite{tHooft:1993dmi,Susskind:1994vu,Maldacena:1997re}. But the map which defines an effective cosmology/'pseudo'-\emph{QFT} correspondence obtained by composition of holography and \emph{DW/C} correspondence, may need to be studied in more details. In fact, the authors of \cite{McFadden:2009fg,McFadden:2010na} start with an inflationary cosmology, then build its dual domain-wall solution which has a dual \emph{QFT}. Finally, the `pseudo'-\emph{QFT}---dual to the starting cosmology---is obtained via analytic continuation of the first \emph{QFT}. Although this approach seems perfectly reasonable and effective, it may also be useful to keep track of the \emph{DW/C} correspondence along all calculations.

Indeed, in this paper, we propose to do a new step towards constructing a consistent cosmology/\emph{QFT} correspondence. We will start by defining what we call a \emph{DW/C}-ready metric like in \cite{hep-th/0602260,hep-th/0610253,hep-th/0609056}. This metric introduces a parameter $\eta$ which shows where the \emph{DW/C} correspondence takes hold. When we keep track of this parameter on the quantum gravity side of holography and study how it arises in the \emph{QFT} dual, we define the holographic dual of the \emph{DW/C} correspondence. In fact, in this framework we are able to see how the analytic continuation at the origin of the \emph{DW/C} correspondence, generates \emph{QFT}s dual to cosmological solutions. This ensures we do not have to perform the analytic continuation ourselves on \emph{QFT}s like it was done in \cite{McFadden:2009fg,McFadden:2010na}. However, in this paper, we only test this framework on the toy model that is the \emph{GPPZ} flow, by computing a scalar one-point function and studying how it is affected by the \emph{DW/C} correspondence.

Therefore, this paper is organized a follows. First, Section~\ref{sec:dwccorrespondence} introduces the domain-wall/cosmology correspondence as done in \cite{hep-th/0602260,hep-th/0610253,hep-th/0609056}. Then, Section~\ref{sec:holoren} gives an account on the holographic renormalization procedure \cite{Henningson:1998ey,hep-th/0002230} adapted to the \emph{DW/C} correspondence and with an application to the \emph{GPPZ} flow \cite{Girardello:1999bd}. Section~\ref{sec:perttheory} develops the gauge-invariant perturbative framework necessary to solve the bulk field equations. In Section~\ref{sec:gppzAns} we use everything we defined and computed in this paper to obtain the scalar one-point function dual to the scalar field present in the \emph{GPPZ} flow. Finally, we discuss the implications of the result on the definition of holography for cosmologies.

    \section{The Domain-wall/Cosmology correspondence} \label{sec:dwccorrespondence}

        This section will serve as an introduction to the Domain-Wall/Cosmology (\emph{DW/C}) correspondence as defined in \cite{hep-th/0602260,hep-th/0610253,hep-th/0609056}. However, we will build on their definition with an important caveat. First, let us give a review of the \emph{DW/C} correspondence.

The $(d+1)$-dimensional metric for a maximally symmetric $d$-dimensional domain-wall is given, in convenient coordinates, by
\begin{equation}
    ds^2 = dr^2 + e^{2A(r)}\left( -\frac{d\hat{t}^2}{1+k \hat{t}^2} + \hat{t}^2 d\Omega_{-}^2\right) ~.
\end{equation}
Here $d\Omega_{-}^2$ is the metric on the $d$-dimensional hyperboloid of unit radius with $SO(1,d-1)$ invariance. We can choose its coordinates such that we get
\begin{equation}
    d\Omega_{-}^2 = d\psi^2 + \sinh^2 \psi d\Omega_{d-2}^2 ~.
\end{equation}
Therefore, in this configuration, our domain-wall is characterised by a function $A(r)$ and a constant $k$ which takes values $0,\pm1$ depending on the transverse space's geometry. The cases are as follows: de Sitter geometry for $k=1$, Anti-de Sitter for $k=1$ and Minkowski when $k=0$. Moreover, in this spacetime lives a scalar field $\Phi(r)$ governed by a scalar potential $V$. The scalar field depends only on the radial coordinate $r$ to preserve the isometries of the transverse slices.

Now consider the analytic continuation of some of the coordinates, \emph{i.e.} the following coordinate transformation
\begin{equation}
    (r,\hat{t},\psi) \rightarrow i(t,\bar{r},\theta) ~.
\end{equation}
Which, once applied to the metric given above, gives the metric for an \emph{FLRW} cosmology
\begin{equation}
    ds^2 = - dt^2 + e^{2A(it)}\left( \frac{d\bar{r}^2}{1 - k \bar{r}^2} + \bar{r}^2 d\Omega_{+}^2\right) ~.
\end{equation}
And now, $d\Omega_{+}^2$ is the metric on the unit radius $d$-sphere with $SO(d)$ invariance and its coordinates are given by those of $d\Omega_{-}^2$ after the analytic continuation, \emph{i.e.}
\begin{equation}
    d\Omega_{+}^2 = d\theta^2 + \sin^2 \theta d\Omega_{d-2}^2 ~.
\end{equation}
This transformation seems to relate domain-wall geometries to cosmologies quite well, if not for a little caveat. Indeed, after the transformation, the scale factor becomes $e^{A(it)}$ which can be troublesome. The authors of \cite{hep-th/0602260,hep-th/0610253,hep-th/0609056} argue that the function $A(it)$ appears as being complex when in fact it is a real function by imposing reality when solving the field equations. However, let us go one step further and study this for the example of an $AdS_{d+1}$ domain-wall. In this case, we have $A(r)=r/L_{AdS}$ where $L_{AdS}$ is the bulk's radius. As such, when we analytically continue the metric then we are left with $A(it)=it/L_{AdS}$ which is evidently not real. This can be remedied via analytic continuation of the radius\footnotemark $L_{AdS} \rightarrow iL_{dS}$. This new scale is dubbed $L_{dS}$ because the cosmology generated this way is a \emph{FLRW} universe with a de Sitter geometry. Later, we will investigate asymptotically-\emph{AdS} spaces, meaning we will have $A(r)=r/L_{AdS} + \cdots$ for which it is possible to extend this argument.

\footnotetext{We want to make it clear that this continuation and the argument given in \cite{hep-th/0602260,hep-th/0610253,hep-th/0609056}, are one and the same. We simply show here how it comes into play for this specific example.}

With this in mind we can write down a \emph{DW/C}-ready metric in which a parameter $\eta=\pm1$ is introduced to keep track of the \emph{DW/C} correspondence. In variables similar to the ones in \cite{hep-th/0602260,hep-th/0610253,hep-th/0609056}, it is given by
\begin{equation}
    ds^2 = \eta (fe^{\alpha\varphi})^2 dr^2 + e^{2\beta\varphi} \left[-\frac{\eta d\hat{t}^2}{1+\eta k\hat{t}^2} + \hat{t}^2 d\Omega_{-\eta}^2 \right] ~.
\end{equation}
A monotonic lapse function $f$ is introduced here through the replacement $dr\rightarrow f(r)e^{\alpha\varphi(r)} dr$ to keep the $r$-reparametrization invariance. $\varphi$ is related to $A$ up to the factor $\beta$ such that $\alpha = \beta d = \sqrt{d/2(d-1)}$. Nevertheless, it is possible to choose the gauge where $f=e^{-\alpha\varphi}$ allowing us to get back the metrics we defined above. In such a gauge, when $\eta=-1$ we retrieve the \emph{FLRW} metric whereas $\eta=1$ gives a domain-wall metric. The Euler-Lagrange equations of a general low-energy Lagrangian density with this metric, are the same as those of the following effective Lagrangian
\begin{equation}
    \mathcal{L} = \frac{1}{2} f^{-1} \left(\varphi^{\prime 2} - |\Phi|^{\prime 2} \right) - fe^{2\alpha\varphi} \left( \eta V - \frac{\eta k}{2\beta^2} e^{-2\beta\varphi} \right) ~.
\end{equation}
The overdot stands for $r$-differentiation. One can realize this Lagrangian is the same whether $\eta=1$ with $V$ and $k$ or $\eta=-1$ with $-V$ and $-k$. This is what defines the domain-wall/cosmology correspondence. To put it simply: \emph{domain-wall solutions with transverse curvature $k$ and potential $V$ are dual to cosmologies described by $-k$ and $-V$}.

Because our primary interest is to study cosmologies in later work, we will prefer to flip the sign of $\eta$, \emph{i.e.} cosmological models are now given when $\eta=1$. As such, the metric we will use from now on is
\begin{equation}
    ds^2 = - \eta dt^2 + e^{2A} \left( \frac{\eta d\bar{r}^2}{1-\eta k \bar{r}^2} + \bar{r}^2 d\Omega_{\eta}\right) ~.
\end{equation}

    \section{Applied to holographic renormalization} \label{sec:holoren}

        In this section we introduce the method of holographic renormalization developed in \cite{Henningson:1998ey,hep-th/0002230}. Then we will apply it to the case of the \emph{GPPZ RG}-flow \cite{Girardello:1999bd} in a similar way to what was done in \cite{hep-th/0105276}. However, in this paper we introduce a twist to those calculations: the \emph{DW/C} correspondence. Unlike previous work, here we will use a \emph{DW/C}-ready metric, as defined in the previous section. But first, we will give some motivations as to why holographic renormalization is needed.

An \emph{RG}-flow can be holographically described by a domain-wall spacetime in which scalar fields live, on the gravitational side of holography. But on the gauge theory side, it represents the running of coupling constants. Take the example of a flow connecting two distinct \emph{CFT}s, then the intermediary field theory is not a \emph{CFT} itself. Now the gravity representation of this flow is a domain-wall spacetime asymptotically \emph{AdS} in two places, called conformal boundaries. Near such a boundary, by choosing the right coordinates, the metric can be brought to the following form
\begin{equation}\label{eq:defMetricEu}
	ds^2= \frac{L^2 d\rho^2}{4\rho^2} + \frac{1}{\rho} g_{ij} dx^{i} dx^{j} ~.
\end{equation}
And $g_{ij}$ admits the following near-boundary expansion
\begin{equation}\label{eq:expmetric}
	g = g_{(0)} + \rho g_{(2)} + \ldots + \rho^{d / 2} \left[g_{(d)}+\sum^{d/2}_{k\geq 1}h_{k(d)}\log^k \rho\right]+\mathcal{O}\left(\rho^{d / 2+1}\right) ~.
\end{equation}
This is achieved by starting from the \emph{DW} metric and taking the coordinate $\rho$ to be such that $\rho=exp(-2r/L)$. While we have $A(r)=r/L + \ldots$ near the boundary, since $A(r)=r/L$ is for the pure \emph{AdS} geometry. As for the scalar field, it also admits a near-boundary expansion which depends on the conformal weight $\Delta$ of its holographic counterpart.
\begin{equation}
	\Phi(x, \rho)=\rho^{\frac{d-\Delta}{2}}\left[\phi_{(0)}+\rho \phi_{(2)}+\cdots+\rho^{\frac{2 \Delta-d}{2}}\left(\phi_{(2 \Delta-d)}+\psi_{(2 \Delta-d)}\log \rho\right) \right]+\mathcal{O}\left(\rho^{\frac{\Delta+2}{2}}\right) ~.
\end{equation}
In all near-boundary expansions, the logarithmic terms only appear if $d$ is even.

The relevant part of the supergravity action in this scenario is given by\footnotemark
\begin{equation}
	S = \int_{\mathcal{M}} \textnormal{d}^{d+1}x \sqrt{G} \left[\frac{1}{4}\mathcal{R} + \frac{1}{2}\partial^{\mu}\Phi \partial_{\mu} \Phi +  V(\Phi) \right] - \frac{1}{2} \int_{\partial\mathcal{M}} \textnormal{d}^{d}x \sqrt{\sigma} K ~,
\end{equation}
where $K$ is the trace of the second fundamental form. Once all near-boundary expansions are plugged in this action then one can see divergences arising. Holographic renormalization now comes in to get rid of those infinities and proceeds as follows. First, we regulate the on-shell action by imposing a cut-off $\rho>\epsilon$. Then the field equations are solved iteratively with the near-boundary expansions, to express the divergences as local functions of the sources $\phi_{(0)}$ and $g_{(0)}$ as $\epsilon\rightarrow0$. Then local counterterms are added to cancel the divergences. The counterterms can be expressed in terms of the sources or of local invariants built with $\Phi(x,\epsilon)$ and the induced metric $\sigma\iij = \frac{1}{\epsilon} g\indices{_{(0)}_i_j}$ which are obtained by reversing the near-boundary expansions.
\footnotetext{The action is given with a Euclidean signature as it is the most common signature used in \emph{AdS/CFT}. Later in this paper, we revert back to a Lorentzian action which will make the \emph{DW/C} correspondence appear.}

All of this apparatus is needed to compute holographic $n$-point functions. In fact, let's take the example of the one-point function of the operator $\mathcal{O}$
\begin{equation}
	\langle \mathcal{O}\rangle = \frac{1}{\sqrt{g_{(0)}}}\frac{\delta S_{sub}}{\delta\phi_{(0)}} = \lim_{\epsilon \to 0} \left( \frac{1}{\epsilon^{\Delta/2}}\frac{1}{\sqrt{\sigma}} \frac{\delta S_{sub}}{\delta\Phi} \right) ~.
\end{equation}
This is given by variation of the action in the GKP-Witten relation \cite{Gubser:1998bc,Witten:1998qj}. $S_{\textnormal{sub}}$ is such that, if $S_{\textnormal{ren}}$ is the renormalized action then $S_{\textnormal{ren}} = \lim_{\epsilon \to 0} S_{\textnormal{sub}}$. However, in the renormalized action, not all divergences can be expressed in terms of sources by only using near-boundary analysis. In fact, for scalar fields, the coefficient $\phi_{(2 \Delta-d)}$ is never determined this way. But this can be remedied by a deep-interior analysis, \emph{i.e.} solve the equations of motion in the interior of the bulk.

Now it is clear why holographic renormalization is needed. Nevertheless, counterterms---and thus scalar $n$-point functions---depend on the exact expression of the scalar potential $V(\Phi)$. So let us apply the procedure we described above to the case of the \emph{GPPZ} flow. All while we keep track of the \emph{DW/C} correspondence as it is the main goal of this paper.

        \subsection{For the GPPZ flow}

            The \emph{GPPZ} \emph{RG}-flow \cite{Girardello:1999bd} goes from a $\mathcal{N}=4$ super Yang-Mills (SYM) theory in the ultraviolet to a pure $\mathcal{N}=1$ SYM in the infrared. According to holography, the UV fixed point---\emph{i.e.} the $\mathcal{N}=4$ SYM---is dual to Type IIB string theory on $AdS_{5}\times S^{5}$. On the IR side, the field theory is dual to a domain-wall solution of five-dimensional gauged supergravity with $\mathcal{N}=1$ supersymmetry. This flow is generated by the insertion of a mass term for the three fermions in the $\mathcal{N}=1$ chiral multiplet of the $\mathcal{N}=4$ SYM. Therefore, this flow has no ties to cosmology, but we use it nonetheless because our goal is to understand how the \emph{DW/C} correspondence changes $n$-point functions. And what better way to assess this change than to use a flow---\emph{i.e.} a domain-wall solution---extensively studied.

The full solution for the \emph{GPPZ} flow is given by the superpotential $\mathcal{W}$, the expressions for $\Phi(\rho)$ and $e^{2A}$. However for the purpose of this section we only require the superpotential which is given by
\begin{equation}
	\mathcal{W}(\Phi) = -\frac{3}{4} \left[ 1 + \cosh\left( \frac{2\Phi}{\sqrt{3}} \right) \right] ~.
\end{equation}
And this allows us to get the scalar potential expression via
\begin{equation}
	V = \frac{\lambda^2}{8} \sum_{\Phi} \left\lvert \frac{\partial \mathcal{W}}{\partial\Phi} \right\rvert^2 - \frac{\lambda^2}{3} \lvert \mathcal{W}\rvert^2 ~,
\end{equation}
with $\lambda=2/L$ in $d=4$ dimensions---which is the case we are interested in. The simplest domain-wall solution have $g\iij=\delta\iij$ in \eqref{eq:defMetricEu} which would allow us to get a simple expression for the regularized on-shell action \cite{hep-th/0105276,Skenderis:1999mm}
\begin{equation}\label{eq:subDivergence}
	S_{\textnormal{reg}} = \int_{\rho=\epsilon}\textnormal{d}^4 x \sqrt{\sigma} \mathcal{W}[\Phi] = \int_{\rho=\epsilon}\textnormal{d}^4 x \sqrt{\sigma} \left( -\frac{3}{2} - \frac{1}{2}\Phi^2 - \frac{1}{18}\Phi^4 + \mathcal{O}(\Phi^5) \right) ~.
\end{equation}
The second equality is obtained by expanding the superpotential around $\Phi=0$, \emph{i.e.} close to the conformal boundary. We can see here that once the near-boundary expansions are plugged in, the first two terms create divergences. This means we expect counterterms to contain at least enough terms to cancel these divergences. On the other hand the term in $\Phi^4$ has a finite limit which means the the on-shell action will not be zero. Indeed, supersymmetry requires the vacuum energy to be zero, \emph{i.e.} the on-shell action has to be zero. Thus we will have to subtract this finite term, which is known as fixing the supersymmetric scheme. We choose this scheme simply because the \emph{GPPZ} flow is supersymmetric.

Nevertheless, because fixing $g\iij=\delta\iij$ is a subclass of solutions of the theory, the counterterms obtained via \eqref{eq:subDivergence} are the most general results we can produce. For this purpose we take $g\iij=g\iij(x,\rho)$. Furthermore, since it is the goal of this paper, we now introduce the \emph{DW/C} correspondence through the metric which now reads
\begin{equation}\label{eq:defGenMetric}
	ds^2=-\eta \frac{L^2 d\rho^2}{4\rho^2} + \frac{1}{\rho} g\iij dx^{i} dx^{j} ~.
\end{equation}
The action also contains markers of the correspondence since it affects the scalar potential,
\begin{equation}\label{eq:fullAction}
	S = \int_{\mathcal{M}} \textnormal{d}^{d+1}x \sqrt{-G} \left[\frac{1}{4}\mathcal{R} - \frac{1}{2}\partial^{\mu}\Phi \partial_{\mu} \Phi - \eta V(\Phi) \right] + \frac{1}{2} \int_{\partial\mathcal{M}} \textnormal{d}^{d}x \sqrt{\eta \sigma} K ~.
\end{equation}
After computing the field equations for this action, one can use them with the near-boundary expansions for the metric and a scalar field with $\Delta=3$, to find a number of relations between the different coefficients in the expansions. An important point that arises from the Klein-Gordon equations for the scalar field is that in this prescription we have
\begin{equation}
	\eta m^2=-\frac{1}{L^2}\Delta(\Delta-d) ~.
\end{equation}
This further highlights the correspondence between a cosmological scalar potential $V$ and its dual potential $-V$ in the domain-wall solution. The procedure goes as follows \cite{Henningson:1998ey,hep-th/0002230}. After iteratively solving the field equations with the expansions, we can regularize the action by imposing the cut-off $\rho>\epsilon$, using the expansions and the relations from the previous step. Once all divergent terms have been expressed in terms of sources---when possible---the near-boundary expansion can be ``inverted'' to give functions such as $\Phi(x,\epsilon)$ or $\mathcal{R}[\sigma]$, with $\sigma\iij=\frac{1}{\epsilon}g_{(0)ij}$. And thus we get the on-shell regularized action
\begin{equation}\label{eq:gppzRegAction}
	\begin{split}
		S_{\textnormal{reg}} &= \frac{\eta}{L} \int_{\partial\mathcal{M}} \textnormal{d}^4 x \sqrt{\eta\sigma} \left( \frac{3}{2} + \frac{1}{2}\Phi^2 + \frac{1}{8}\eta \mathcal{R} \right. \\ 
		& \quad \left. + \log\epsilon \left[ \frac{L^4}{32}( \frac{1}{3}\mathcal{R}^2 - \mathcal{R}^{ij}\mathcal{R}_{ij}) + \frac{\eta}{4}(\Phi\square_{\sigma}\Phi + \frac{1}{6}\mathcal{R}\Phi^2) \right] + \ldots \right) ~.
	\end{split}
\end{equation}
Here we kept only the terms that generate divergences and, in there, we can find the terms already present in \eqref{eq:subDivergence}. From this result we can see that there is a non-trivial dependence on the \emph{DW/C} correspondence. However, we cannot draw conclusions this early, as the quantity we are trying to study is the one-point function $\langle \mathcal{O} \rangle$. To compute it we still need to add the counterterms to the regularized on-shell action which we can deduce from the equation above. Thus we get the variation of the action with respect to the scalar field $\Phi$
\begin{equation}\label{eq:varAction}
	\begin{split}
			\delta S_{\textnormal{sub}} &= \delta S_{\textnormal{reg}} + \delta S_{\textnormal{ct}} + \delta S_{\textnormal{ct,fin}} \\
			&=  \int_{\mathcal{M}} \textnormal{d}^5 x \sqrt{-G} \left[ \square_{G}\Phi - \eta V' \right]\delta\Phi \\
			&\quad - \frac{\eta}{L} \int_{\rho=\epsilon}\textnormal{d}^4 x \sqrt{\eta\sigma} \left[ - 2\epsilon\partial\indices{_\epsilon}\Phi + \Phi + \frac{\eta}{2}\log\epsilon (\square_{\sigma}\Phi + \frac{1}{6}\mathcal{R}\Phi) + \frac{2}{9}\Phi^3 \right] \delta\Phi ~.
		\end{split}
\end{equation}
As mentioned before, we fix the supersymmetric scheme by introducing a finite counterterm $S_{\textnormal{ct,fin}}$. In equation \eqref{eq:varAction}, the first integral will vanish since it is evaluated on-shell. This leaves us with the second integral which will give the vacuum expectation value we are searching for,
\begin{equation}\label{eq:vev1st}
	\langle \mathcal{O}\rangle = \lim_{\epsilon \to 0} \left( \frac{1}{\epsilon^{\Delta/2}}\frac{1}{\sqrt{\eta\sigma}} \frac{\delta S_{\textnormal{sub}}}{\delta\Phi} \right)
		= \frac{\eta}{L} \left[ 2( \psi_{(2)} + \phi_{(2)} ) - \frac{2}{3} \phi_{(0)}^3 \right] ~.
\end{equation}
First, we notice the presence of $\eta$ which means the \emph{DW/C} correspondence reaches this result. As a check, we can see that by fixing $\eta=-1$ we retrieve the equation found in \cite{hep-th/0105276}. Unfortunately this equation is not exactly the result sought after. In fact, earlier in this section, we mentioned that coefficients like $\phi_{(2)}$ cannot be expressed as functions of the sources by using only near-boundary analysis. Actually, a solution to the bulk field equations is required to find $\phi_{(2)}$. However, a solution to the nonlinear equations is too difficult to achieve. On the other hand, linearizing the equations is the way to go. And this is exactly what we do in the next section.

    \section{Consolidated with cosmological perturbation theory} \label{sec:perttheory}

        %Mention this: in this section we use the formalism developed by Mukhanov for cosmological perturbation theory because although it possible to define gauge-invariant variable like Skenderis does it, using Mukhanov adds the possibility for a cosmological interpretation of equations when looking at the cosmology side of the correspondence
%
The purpose of this section is to introduce the framework we use to linerize the bulk field equations. We will use the theory of cosmological perturbation as developed by Mukhanov in \cite{Mukhanov} which focuses on a gauge-invariant linearization. This approach differs from the one taken in \cite{hep-th/0105276} mainly in our relation to the definition of gauge-invariant variables. Indeed, after the publication of \cite{DeWolfe:2000xi}, it was noted by the authors of \cite{hep-th/0105276} that a gauge-invariant treatment of the bulk field equations is primordial to ensure the generality of the results we need. Thus they introduced pragmatic gauge-invariant variables serving the sole purpose of solving the equations somewhat easily. Thanks to cosmological perturbation theory, we add the possibility of using the variables we introduce to be further employed to study cosmologies generated by the \emph{DW/C} correspondence. However, it is important to note the \emph{DW/C} correspondence does not require cosmological perturbation theory to function with the holographic framework. And vice-versa, this perturbative framework can be applied to generic holography, as it was shown in \cite{vaduret2019gppz}.

Let us begin our study of cosmological perturbation theory by stating that we are interested only in scalar perturbations, because we strive to find scalar one-point functions. Moreover, any other type of perturbations decouple from the one we chose. As such, the perturbed \emph{DW/C}-ready metric is given by
\begin{equation}\label{eq:defPertElement}
	ds^2 = -\eta(1+2\chi)dr^2 + e^{2A(r)}\left[ 2\partiali E dx\indices{^i}dr + \left[(1+2\psi)\Sigma\indices{_i_j} - 2\partiali\partialj W\right]dx\indices{^i}dx\indices{^j} \right] ~,
\end{equation}
with $\Sigma$ the diagonal metric $\Sigma\iij=\operatorname{diag}(\eta,1,\cdots,1)$, and the scalar perturbations $\chi(x,r)$, $E(x,r)$, $\psi(x,r)$ and $W(x,r)$. We also consider perturbations of the scalar field
\begin{equation}\label{eq:defPertPhi}
	\Phi(r,x\indices{^i}) = \bkg{\Phi}(r) + \prt{\Phi}(r,x^{i}) ~.
\end{equation}
Note that the superscripts $\bkg{\Phi}$ refers to background values while $\prt{\Phi}$ refers to first-order perturbations. By definition, these perturbations are all considered infinitesimal. Moreover, they all admit a near-boundary analysis which we will specify when needed. Inspired by cosmologists' notations, we call $a=e^{A(r)}$ the scale factor and we define the Hubble parameter $H=a'/a=A'$.

Even after choosing this coordinate system there remain some gauge freedom. In fact, one can still perform one-parameter infinitesimal gauge transformations such that
\begin{equation}\label{eq:gaugeTrans}
	x\indices{^\mu} \rightarrow x\indices{^\mu} + \xi\indices{^\mu}(r,x\indices{^i}) ~,
\end{equation}
where $\xi\indices{^\mu}=(\xi\indices{^r}(x,r),~\xi\indices{^i})$ drives the transformation and is comprised of a function $\xi\indices{^r}(x,r)$ and a vector $\xi\indices{^i}$. The latter is such that
\begin{equation}
	\xi\indices{^i} = \xi\indices{^i_{\textnormal{tr}}} + \Sigma\indices{^i^j}\nabla\indices{_j}\xi(x,r) ~,
\end{equation}
with $\nabla\indices{_i}\xi\indices{^i_{\textnormal{tr}}} = 0$ and a function $\xi(x,r)$. These conditions ensure that we only work with scalar gauge transformations. Now, all the perturbation fields transform under \eqref{eq:gaugeTrans} and the field equations as well. The expression of those transformations can be found in Appendix~\ref{sec:appendix}. After studying how those quantities transform, one can reverse-engineer new variables which would be invariant under \eqref{eq:gaugeTrans}. We call these gauge-invariant variables and the simplest we can make up are the following
\begin{equation}\label{eq:giVar}
	\begin{gathered}
		R = \chi + \eta \left( a^2 (E + W') \right)' ~; \\
		\Psi = \psi + \eta a a' (E + W') ~; \\
		\gif{\Phi} = \prt{\Phi} - a^2 \eta (E + W') \bkg{\Phi}' ~.
	\end{gathered}
\end{equation}
The new superscript $\gif{\Phi}$ refers to gauge-invariance, and thus inly appears next to first-order perturbations. Primes refer to differentiation with respect to $r$. The establishment of these variables also means we can create gauge-invariant field equations, after studying how they transform. Both of these can also be found in Appendix~\ref{sec:appendix}. In fact, we use the gauge-invariant Einstein equations to the equation of motion for one of $R$, $\Psi$ or $\gif{\Phi}$. The first two are found to be proportional to each other, while they are related to $\gif{\Phi}$ via the $(ri)$ Einstein equation \eqref{eq:GIEin}.

The reader may remember we need to solve the deep-interior linearized field equations in order to find missing expressions for some near-boundary expansion coefficients. After playing around with the gauge-invariant Einstein equations, we can find the EOM for $R$. And it is given by
\begin{equation}\label{eq:eomR}
	R'' + \left[ (d-2)H - \frac{H''}{H'} \right] R' + (d-2)\left[ 2H' - \frac{HH''}{H'} + \eta\frac{1}{a^{2}(d-2)}p^2 \right] R = 0 ~.
\end{equation}
Note that we went from position space to momentum space by introducing the transverse momentum $p^2$. Although we defined gauge-invariant variables differently than what was done in \cite{hep-th/0105276}, it is possible to convert \eqref{eq:eomR} into their equation $(4.13)$. Indeed, we can find that their variable $R$---which we rename $R_S$---is such that $p^2 R = H' a^2 R_S$ for domain-wall solutions. The attentive reader will also have noticed the apparition of $\eta$ in the equation above. This entails that it may also appear in the solution.

However, finding a solution to the equation without an expression for $A(r)$ seems too tedious and difficult. So instead, let us pick up where we left off and apply what we just found to the example of the \emph{GPPZ} flow. In which case, if \eqref{eq:eomR} admits a solution then we will be able to use it to finally compute one-point functions.

    \section{For the GPPZ flow} \label{sec:gppzAns}

        With the knowledge from the previous sections, we are now ready to compute the scalar one-point function for the \emph{GPPZ} flow. First, the rest of the \emph{GPPZ} domain-wall solution is given by
\begin{equation}\label{eq:bkgSol}
	\bkg{\Phi} = \frac{\sqrt{3}}{2}\log\frac{1 + \sqrt{\rho}}{1 - \sqrt{\rho}}~, \quad e^{2A} = \frac{1-\rho}{\rho} ~.
\end{equation}
With this and the \emph{DW/C} correspondence, we can also take a look at the cosmological model generated by this \emph{RG}-flow. In fact, when we compute the effective Friedmann equations, we find a $5$-dimensional universe containing radiation, spatial curvature and vacuum densities.

This gives us everything we need to find a solution to \eqref{eq:eomR}. In fact, the equation is now
\begin{equation}
	(1-\rho)R'' + \frac{1}{\rho} \left[ -(1+2\rho) R' + \left( \frac{1}{\rho} + \frac{\eta L^2 p^2}{4} \right)R \right] = 0 ~.
\end{equation}
$R$ represents infinitesimal perturbations, and so we expect its expression to behave as such, \emph{i.e.} it should not blow up anywhere in the bulk. However, the equation above has two solutions, one of which goes to infinity in the deep interior of the bulk. Thus we keep only the following
\begin{equation}\label{eq:solR}
	R(p^2 , \rho) = \rho F\left(\frac{3}{2}+\frac{\sqrt{1+ \eta p^2L^2}}{2},~ \frac{3}{2}-\frac{\sqrt{1+ \eta p^2L^2}}{2};~ 3;~ 1-\rho\right) ~,
\end{equation}
since the field equations are valid all over the bulk, and so is this solution. This means we can use it on the boundary to find the expression for $\phi_{(2)}$ we are missing. Hypergeometric functions can be expressed as series for certain values of its parameters. Fortunately enough, the function given above has such an expansion which can be found in \cite{bateman1981higher}.

We mentioned in the previous section that the metric perturbation fields also had a near-boundary expansion. The only one we need in this paper is
\begin{equation}\label{eq:expPW}
	\psi = \frac{1}{1-\rho} \left[h_{(0)}+h_{(2)}\rho+\rho^2(h_{(4)}+hh_{1(4)}\log\rho+hh_{2(4)}\log^2\rho)+...\right] ~.
\end{equation}
The prefactor $1/(1-\rho)$ is introduced to keep a straightforward correspondence between the coefficients of \eqref{eq:expPW} and the ones in the expansion of $g\iij$ \eqref{eq:expmetric}. $E$, $W$ and $\chi$ admit similar expansions. We do not need to specify the other expansions because we fix the gauge such that $\chi=E=0$ and we will not present intermediary steps in which $W$ appears, simply because they are straightforward and tedious. Nevertheless, here is a run down of what to expect. First, one can use the definition of the gauge-invariant variables \eqref{eq:giVar}, the relation between $R$ and $\Psi$ and \eqref{eq:solR} to find useful relations concerning coefficients of the near-boundary expansions of $W$ and $\psi$. These relations can then be plugged in the $ri$ component of the gauge-invariant Einstein equation, along with necessary expansions, to find a number of expressions which once combined give
\begin{equation}\label{eq:phi2psi2}
	\phi_{(2)} + \psi_{(2)} = \phi_{(0)} + \eta \frac{L^2 p^2}{4} (\phi_{(0)} + \sqrt{3} h_{(0)}) \alpha_{0} ~.
\end{equation}
With this we are now able to complete \eqref{eq:vev1st}. However, we need to linearize \eqref{eq:vev1st} since we are dealing with perturbations. This can be done on-the-fly and it only affects its last term to become $-\frac{2}{3}\prt{\phi}_{(0)}\bkg{\phi}_{(0)}^2$ while \eqref{eq:bkgSol} gives us $\bkg{\phi}_{(0)}=\sqrt{3}$. This gives us the long anticipated result
\begin{equation}\label{eq:finalResult}
	\langle \mathcal{O}\rangle = \frac{L p^2}{2} (\prt{\phi}_{(0)} + \sqrt{3} h_{(0)}) \alpha_{0} ~.
\end{equation}
This one-point function starts the holographic description of a scalar field generating a cosmology with radiation, spatial curvature and vacuum densities but devoid of ordinary matter. On a side note, one can see from solution \eqref{eq:solR} that the parameter $p^2$ can take any value.

The immediate observation one can make about this result is that there is no $\eta$. This evidently means the \emph{DW/C} correspondence has no reach over this one-point function. But this raises the question: why is that? We saw all along the calculations that $\eta$ appears at every steps but it is not here anyway. The first justification that comes to mind, is the presence of an underlying symmetry. To assess whether this symmetry exists or not, we need to compute other \emph{CFT}-related quantities. Correlators, for example, may give us some better insight in this endeavour. Appendix A of \cite{hep-th/0209067} gives some hints as to what would happen to the energy-momentum tensor when analytically continuing \emph{AdS} to \emph{dS} spaces. However, their continuation may not contain subtleties that the \emph{DW/C} correspondence does. Instead of a symmetry, the simplest approach may be to argue that under the \emph{DW/C} correspondence, some domain-wall ``equals'' to some cosmology thus these are equivalent to one and only one \emph{CFT} through holography. However this may seem far-fetched since the \emph{CFT} dual to the domain-wall solution is Lorentzian whereas the one dual to the cosmology is Euclidean.

Moreover, one will have noticed the overall factor $L$. Because of it, if one tries to perform the full analytic continuation responsible of the \emph{DW/C} correspondence on this expression\footnotemark, then \eqref{eq:finalResult} picks up a factor $i$. If we think about it for a second, we realise that under the \emph{DW/C} correspondence, the boundary goes from Lorentzian to Euclidean and vice-versa. We also know that this move makes a factor of $i$ appear in the generating functional of the \emph{CFT}. Thus this $L$ carries the $i$ factor from the Lorentzian-to-Euclidean continuation. This shows that the \emph{DW/C} correspondence is indeed dual to an analytic continuation between Lorentzian and Euclidean \emph{QFT}s, as suggested in \cite{McFadden:2009fg,McFadden:2010na}. Equation \eqref{eq:finalResult} does not contain this imaginary factor since the Lorentzian-to-Euclidean continuation for the \emph{CFT} is assumed when fixing $\eta$.

\footnotetext{This has to be done instead of keeping track of $\eta$ like we did, since these methods should be equivalent.}

    \section{Conclusions}

        In this paper we developed a consistent way to apply the domain-wall/cosmology correspondence to the \emph{AdS/CFT} correspondence. We kept track of the marker of the \emph{DW/C} correspondence during the calculations. Instead of using pragmatic perturbation theory, we introduced cosmological perturbations theory by Mukhanov \cite{Mukhanov}, which may reveal itself to be useful to study the cosmology side of the \emph{DW/C} correspondence. Fortunately this does not lengthen calculations. Finally we were able to get a scalar one-point function for the example of the \emph{GPPZ} flow. And we found that this results is not dependent on the \emph{DW/C} correspondence. This may highlight a symmetry, but the results we provide are not enough to make any assertions yet. On the other hand, we showed with this result that the \emph{DW/C} correspondence is indeed dual to an analytic continuation, like it was introduced by hand in \cite{McFadden:2009fg,McFadden:2010na}. 

However, there is still much to do. A first task will be to compute $n$-point functions to reveal if they are invariant under the \emph{DW/C} correspondence as well. Then a rather straightforward application of the framework we presented here would be to adapt it to 4D cosmologies and study their holographic duals. This would allow us to be truly on par with the work done in \cite{McFadden:2009fg,McFadden:2010na}, to compare our approach to theirs. Moreover, with recent attempts to find holographic descriptions to puzzles of hot Big Bang cosmology \cite{Nastase:2019rsn,McFadden:2013ria}, and efforts to test these conjectures against observational data \cite{Afshordi:2017ihr,Afshordi:2016dvb}, it would be extremely interesting to see how holographic observables depend on the $\eta$-parameter.

Finally, a new braneworld scenario was recently proposed in \cite{Banerjee:2019fzz,Banerjee:2018qey,Banerjee:2020wov}, where a de Sitter cosmology emerges on the interface between two \emph{AdS} vacua. Given a false $AdS_5$ vacuum which decays into a true $AdS_5$ vacuum with a smaller cosmological constant, the brane at the interface of these vacua expands and harbours a de Sitter cosmology. In this set-up, one can easily convince oneself that, through the bubble's expansion, there is a direct correspondence between the radial coordinate---along which the bubble expands---and time on the bubble. This begs the question of the connection between this braneworld scenario and the domain-wall/cosmology correspondence. In light of the framework and result we presented in this paper, the author aims at investigating this mystery in upcoming work.

    \section*{Acknowledgements}
    JFV would like to thank Ulf Danielsson for his kindness and help in writing this paper. Special thanks as well to Marjorie Schillo for supervising the master's thesis from which this paper is extracted. Finally, thank you to Roman Mauch and Daniel Panizo for the helpful discussions.

    \appendix
        \section{Results of cosmological perturbation theory} \label{sec:appendix}
            This appendix gives a collection of results from cosmological perturbation theory with a \emph{DW/C}-ready metric. First, under the gauge transformation \eqref{eq:gaugeTrans} the metric transforms as
\begin{equation}
	G\imunu \rightarrow G\imunu - \nabla\indices{_\mu}\xi\indices{_\nu}(r,x\indices{^i}) - \nabla\indices{_\nu}\xi\indices{_\mu}(r,x\indices{^i}) ~.
\end{equation}
This gives the transformation for the perturbation fields
\begin{equation}\label{eq:gaugeTransFields}
	\begin{split}
		\chi &\rightarrow \chi - \partial\indices{_r}\xi\indices{^r} ~;\\
		\psi &\rightarrow \psi - H\xi\indices{^r} ~;\\
		W &\rightarrow W + \xi ~;\\
		E &\rightarrow E + \eta\frac{1}{a^2}\xi\indices{^r} -\partial\indices{_r}\xi ~.
	\end{split}
\end{equation}
The tranformation of the scalar field perturbation is found by using a Taylor expansion, thus
\begin{equation}
	\prt{\Phi} \rightarrow \prt{\Phi} - \bkg{\Phi}' \xi\indices{^r} ~.
\end{equation}
Then, under a coordinate change $x\indices{^\mu} \rightarrow \tilde{x}\indices{^\mu}$ any tensor $\mathcal{T}$ transforms as
\begin{equation}\label{eq:tensorTrans}
		\mathcal{T}\imunu \rightarrow \widetilde{\mathcal{T}}\imunu(\tilde{x}) = \frac{\partial x\indices{^\sigma}}{\partial \tilde{x}\indices{^\mu}} \frac{\partial x\indices{^\rho}}{\partial \tilde{x}\indices{^\nu}} \mathcal{T}\indices{_\sigma_\rho}(x) ~.
\end{equation}
This means the perturbation part of the Einstein tensor (or energy-momentum tensor) transforms under \eqref{eq:gaugeTrans} as
\begin{equation}
	\begin{split}
		\prtt{\mathcal{G}}{_i_j} &\rightarrow \prtt{\mathcal{G}}{_i_j} - \bkgt{\mathcal{G}}{_m_j} \partiali\xi\indices{^m} - \bkgt{\mathcal{G}}{_n_i} \partialj\xi\indices{^n} - (\bkgt{\mathcal{G}}{_i_j})'\xi\indices{^r} ~,\\
		\prtt{\mathcal{G}}{_r_r} &\rightarrow \prtt{\mathcal{G}}{_r_r} - 2 \bkgt{\mathcal{G}}{_r_r} \partial\indices{_r} \xi\indices{^r} - (\bkgt{\mathcal{G}}{_r_r})'\xi\indices{^r} ~,\\
		\prtt{\mathcal{G}}{_r_i} &\rightarrow \prtt{\mathcal{G}}{_r_i} - \bkgt{\mathcal{G}}{_i_i}\partial\indices{_r}\partiali\xi - \bkgt{\mathcal{G}}{_r_r}\partiali\xi\indices{^r} ~.
	\end{split}
\end{equation}
Now, here are the tensors making up the perturbative Einstein field equations.
\begin{equation}
	\begin{split}
		\prtt{\mathcal{G}}{_r_r} &= (d-1)\left[ dH\psi' - a^{-2} \partialui\partiali (\eta\psi + aa'( E + W')) \right] ~; \\[2ex]
		\prtt{\mathcal{G}}{_i_j} &= \left\{ a^2 [dH^2 + 2H'] (d-1) (\chi - \psi) + (d-1)aa'\chi' \right. \\
		&\quad - \left. d(d-1)aa'\psi' - (d-1) a^2 \psi'' + \eta\partial\indices{^k}\partial\indices{_k}\Omega \right\}\eta\Sigma\indices{_i_j} \\
		&\quad - \partiali\partialj \left\{ \Omega - \eta (d-1) a^2 [dH^2 + 2H'] W \right\} ~; \\[2ex]
		\prtt{\mathcal{G}}{_r_i} &= \partiali\left\{ (d-1)(H\chi-\psi') - \eta E (d-1) a^2 \left( \frac{d}{2} H^2 + H' \right) \right\} ~; \\[2ex]
		\prtt{T}{_r_r} &= 2\chi V + \prt{\Phi} V' + \prt{\Phi}'\bkg{\Phi}' ~; \\[2ex]
		\quad \prtt{T}{_r_i} &= \bkg{\Phi}'\partiali\prt{\Phi} + \left[ \frac{1}{2} \bkg{\Phi}'^2 - V \right] a^2 \eta \partiali E ~; \\[2ex]
		\prtt{T}{_i_j} &= a^2 \eta \Sigma\indices{_i_j} [ \prt{\Phi}'\bkg{\Phi}' - \chi\bkg{\Phi}'^2 - \prt{\Phi} V'] \\
		&\quad + 2a^2 \eta ( \psi\Sigma\indices{_i_j} - \partiali\partialj W) \left[ \frac{1}{2} \bkg{\Phi}'^2 - V \right] ~.
	\end{split}
\end{equation}
And their gauge-invariant versions.
\begin{equation}\label{eq:GIEin}
	\begin{split}
		\git{\mathcal{G}}{_i_j} &= \left\{ a^2 (d-1)[dH^2 + 2H'](R-\Psi) + aa'(d-1)(R'-d\Psi') \right. \\ 
		& \quad \left. - a^2 (d-1)\Psi'' + \eta\nabla^2 (R + (d-2)\Psi) \right\}\eta\Sigma\indices{_i_j} - \partiali\partialj [R + (d-2)\Psi] ~;\\[2ex]
		\git{\mathcal{G}}{_r_r} &= \frac{d-1}{a^2} \left[ daa'\Psi' - \eta\nabla^2\Psi \right] ~;\\[2ex]
		\git{\mathcal{G}}{_r_i} &= (d-1)\partiali\left[ HR-\Psi'\right] ~;\\[2ex]
		\git{T}{_i_j} &= a^2 \left[ \gif{\Phi}'\bkg{\Phi}' - R (\bkg{\Phi}')^2 + \eta \gif{\Phi} V' + 2\Psi (\frac{1}{2}(\bkg{\Phi}')^2 + \eta V) \right]\eta\Sigma\indices{_i_j} ~;\\[2ex]
		\git{T}{_r_r} &= -2\eta R V + \gif{\Phi}'\bkg{\Phi}' - \eta\gif{\Phi} V' ~;\\[2ex]
		\git{T}{_r_i} &= \bkg{\Phi}'\partiali\gif{\Phi} ~.
	\end{split}
\end{equation}

    \bibliographystyle{JHEP}
    \bibliography{archive}

\end{document}